\documentstyle[12pt]{article}
\begin{document}

\title{The string solution in SU(2) Yang-Mills-Higgs theory}

\author{V.D.Dzhunushaliev
\thanks{E-mail: dzhun@freenet.bishkek.su}
and A.A.Fomin}

\date{}

\maketitle

\begin{center}
Theoretical physics department, the Kyrgyz State National
University, 720024, Bishkek, Kyrgyzstan
\end{center}

\begin{abstract}
The tube solutions in Yang - Mills - Higgs theory are received, in
which the Higgs field has the negative energy  density.  This
solutions make up the discrete spectrum numered by  two  integer
and have the  finite  linear  energy  density. Ignoring  its
transverse size, such field  configuration  is the rest infinity
straight string.
\end{abstract}

\centerline{PACS number: 03.65.Pm; 11.17.-w}

\bigskip
\par
At the end of 50-th years W.Heisenberg has been investigate
the nonlinear spinor matter theory (see, for example,
\cite {ivan}, \cite{fin}). It is supposed that on the basis
one or another nonlinear
spinor equation the basic parameters of the elementary particles
existing at that time will be derived: masses, charges and so on.
The mathematical essence of this theory lies in the fact that the nonlinear
spinor Heisenberg equation (HE) (or in the simpler case the nonlinear
boson equation like nonlinear Schr\"odinger equation) has the discrete
spectrum of the solutions having physical meaning (possesing, for example,
the finite energy). This solutions give the mass spectrum in classical
region even. This gave hope that after quantization more or less
likely mass spectrum and the charges of the elementary
particles would be derive.
\par
Now the string can to arise in Dirac theory with the massive vector
field $A_\mu$ by interaction 2 magnetic charges with opposite sign
\cite{nam}. At present time the investigations continue
along this line and explore not 1-dimensional object (string)
stretched between quarks (see, for example, \cite{bars}) but 3-dimensional
(tube) filled by field (see, for example, \cite{nus}, \cite{ols}). So, for
example, a tube of the chromodynamical field and its properties in
\cite{nus} is considered. But this consideration is phenomenological
because a question on the reason of the field pinching isn't affected,
also a question on the field distribution in the tube isn't analyzed.
\par
In this article we shown that the Yang - Mills field interacted
with Higgs scalar field is confined in tube.
In this case the Higgs field have the negative energy density.
\par
In \cite{fin} it is showed that the nonlinear Klein - Gordon  and
Heisenberg equations have the regular solutions. They are the
spherical symmetric particlelike solutions numered by integer,
i.e. they form discrete spectrum with the corresponding energy value.
One would expect (and this will be showed below) that we have in
axial symmetric case as well as in spherical - symmetric case
the physical interesting (string) solutions with finite energy per
unith length.
\par
Finally, we present some qualitative argument in favour of the
existence such field configurations (tube, string) according \cite{bars}.
In QCD vacuum field taken external pressure on the gluon
tube. Diameter of such tube will be defined from equilibrium condition
between external pressure of the vacuum field and internal pressure of
the gluon field in tube. It can be evaluate by minimizing the energy
density of such tube which is the difference between the positive
energy density of the chromodynamic field and negative energy density
of vacuum field in QCD. This diameter $R_0$ after corresponding
calculations is equal:
\begin{equation}
R_0 = \frac{\Phi}{\left ( 2 \pi ^2B\right )^{1/4}},
\end{equation}
where $\Phi$ is a gluon field flux generating quark - antiquark pair,
$(-B)$ is the negative energy density of the vacuum chromodynamical
field.
\par
We seek a self-consistent solution of the Yang - Mills - Higgs
equations system described the Yang - Mills field
confined in the tube. We choose Lagrangian in the following form:
\begin{equation}
{\cal L} = -\frac{1}{4g^2} F_{a\mu\nu} F_a^{\mu\nu} -
\frac{1}{2} \left (D_{\mu} \Phi \right ) \left (
D^{\mu} \Phi \right ) - V(\Phi),
\label{2.1}
\end{equation}
where $a=1,2,3$ is $SU(2)$ colour index; $\mu ,\nu = 0,1,2,3$ are spacetime
indexes;
$F_{a\mu\nu} = \partial _{\mu}  A_{a\mu} - \partial _{\nu} A_{a\mu} +
\epsilon _{abc} A_{b\mu}A_{a\nu}$
is the strength tensor of the $SU(2)$ gauge field;
$F_{\mu\nu} = F_{a\mu\nu} t^a, t^a$
are generators of the $SU(2)$ gauge group;
$D_{\mu}\Phi = (\partial _{\mu} + A_{\mu})\Phi$;
$V(\Phi) = \lambda (\Phi ^+ \Phi - 4\eta ^2)/32$;
$g, \eta ,\lambda$ are constant;
$\Phi$ is an isodoublet of the Higgs scalar field;
\par
The Yang - Mills - Higgs equations system look by following
form in this model:
\begin{equation}
\begin{array}{r}
D_{\mu} F_a^{\mu\nu} = (-\gamma)^{-1/2}
\partial _{\mu}\left [(-\gamma)^{1/2} F_a^{\mu\nu}
\right ] + \epsilon_{abc} A_{b\mu} F_c^{\mu\nu} = \\
\frac {g^2}{2} \left \{\left ( D^{\nu} \Phi \right )^+
\left (t^a \Phi \right ) +
\left (t^a\Phi\right )^+\left (D^{\nu}\Phi \right ) \right \},
\end{array}
\label{2.2.1}
\end{equation}
\begin{equation}
D_{\mu}D^{\mu} \Phi = 2 \frac{\partial V(\Phi )}{\partial \Phi ^+},
\label{2.2.2}
\end{equation}
where $\gamma$ is the metrical tensor determinant.
\par
We seek the string solution in the following form: the gauge potential
$A_{a\mu}$ and the isodoublet of the scalar field $\Phi$ we chosen
in cylindrical coordinate system $(z, r, \theta )$ as :
\begin{eqnarray}
A_{1t} & = & 2\eta f(r),
\label{2.3.1}\\
A_{2z} & = & 2\eta v(r),
\label{2.3.2}\\
A_{3\theta} & = & 2\eta r w(r),
\label{2.3.3}\\
\Phi & =& \left (
\begin{array}{c}
2\eta\varphi(r)\\0
\end{array}
\right )
\label{2.3.4}
\end{eqnarray}
\par
By substituting Eq's(\ref{2.3.1}-\ref{2.3.4}) in Eq's(\ref{2.2.1}-
\ref{2.2.2}) we receive the following equations system:
\begin{eqnarray}
f'' + \frac{f'}{x} & = & f \left [4\left ( v^2 + w^2 \right ) -
g^2\varphi ^2 \right ],
\label{2.4.1}\\
v'' + \frac{v'}{x} & = & v \left [4\left ( -f^2 + w^2 \right ) -
g^2\varphi ^2 \right ],
\label{2.4.2}\\
w'' + \frac{w'}{x} - \frac{w}{x^2} & = & w \left [4\left ( -f^2 + v^2 \right ) -
g^2\varphi ^2 \right ],
\label{2.4.3}\\
\varphi '' + \frac{\varphi '}{x} & = &
\varphi \left (-f^2 + v^2 + w^2 +1 -\varphi ^2 \right ),
\label{2.4.4}
\end{eqnarray}
here the dimensionless coordinate $x=r\eta \sqrt {\lambda}$ is
introduced; $(')$ means the derivative with respect to $x$; and the following
renaming are made:
$g^2\lambda ^{-1/2}\to g^2,
f(x)\lambda ^{-1/2}\to f(x),
v(x)\lambda ^{-1/2}\to v(x),
w(x)\lambda ^{-1/2}\to w(x)$.
We will study this system by the numerical tools. In this
article we investigate the easiest case $v=f=0$. Thus system
(\ref{2.4.1}-\ref{2.4.4}) look as:
\begin{eqnarray}
w'' + \frac{w'}{x} -\frac{w}{x^2} & = & -g^2 w \varphi ^2,
\label{2.5.1}\\
\varphi '' + \frac{\varphi '}{x} & = & \varphi
\left (1 + w^2 - \varphi ^2 \right ).
\label{2.5.2}
\end{eqnarray}
\par
We begin the numerical integration of the assumed equations from point
$x=\Delta \ll 1$. For which purpose we expand functions $f$ and $\varphi$
into a series:
\begin{eqnarray}
\varphi & = & \varphi _0 + \varphi _2 \frac{x^2}{2} + \cdots,
\label{2.6.1}\\
w & = & w_0x + w_3 \frac{x^3}{3!} + \cdots ,
\label{2.6.2}
\end{eqnarray}
after which, substituting in Eq's(\ref{2.5.1}-\ref{2.5.2}) we find that:
\begin{eqnarray}
\varphi _2 & = & \frac{1}{2} \varphi _0
\left ( 1 - \varphi ^2_0 \right ),
\label{2.7.1}\\
w_3 & = & -\frac{3}{4} g^2 w_1 \varphi _0^2.
\label{2.7.2}
\end{eqnarray}
We denote the boundary value $\varphi _0 = \varphi ^*$. Then, the numerical
investigation shows that there are regions of the boundary values
$\varphi ^*$ for which either
$\varphi (x) \stackrel{x\to \infty}{\longrightarrow}+1$
or
$\varphi (x) \stackrel{x\to \infty}{\longrightarrow}-1$.
It is sufficiently evident that between two this regions there is a
boundary value $\varphi _n^*(x)$ caused to the exceptional solution
$\varphi _n(x)$ (integer $n$ is the knot number of the function
$\varphi _n(x)$). This solutions is separatrix in phase space
$(\varphi ',\varphi)$.
\par
It can to find by succesive approximation method this exceptional
solution, wich drop to zero as an exponent by $x \to \infty$. We notice
that in \cite{dzh1} the Eq.(\ref{2.4.4}) with $f=v=w=0$ has been
investigated
and the result is analogical achieved her. The spherical symmetric
solution of such equation has been investigated in \cite{fin} and also
was received that only with the discrete boundary value of the
wavefunction a particlelike solutions be exist with the finite energy.
\par
The right side of Eq.(\ref{2.5.1}) tend to zero (by $x\to \infty$)
by this $\varphi _n^*$ value and hence the asymptotical behaviour of
$w(x)$ function may be following:
\begin{eqnarray}
w(x) \approx \pm x,
\label{2.8.1}\\
w(x) \approx \frac{1}{x}.
\label{2.8.2}
\end{eqnarray}
\par
The numerical investigation show that there are the regions of the
parameter $g$ for which
$w'(x) > 0$ $(w(x)\approx + x)$
and there are the regions of the parameter $g$ for which
$w'(x) < 0$ $(w(x)\approx - x)$.
It is sufficiently evident that there is the exceptional value $g^*$
on the boundary between two such regions caused to the exceptional
solution of the system (\ref{2.5.1} - \ref{2.5.2}) such that
$w(x) \approx 1/x$. Analogously it can to find this exceptional solution
$w(x)$ by succesive approximation method. The numerical calculation
show that the $g^*$ value may be enumerate by integer $m$ indexed
the  knot number of the $w(x)$ function.
\par
It can to show that the asymptotical behaviour of $\varphi (x)$ and $w(x)$
functions look as following:
\begin{eqnarray}
\varphi _{mn} (x) &\approx &
\frac{\exp \left (-x \right )}
{\sqrt{x}},
\label{2.9.1}\\
w_{mn} (x) &\approx &\frac{C}{x} - \frac{Cg^2}{4}\frac{\exp{(-2x)}}{x^2},
\label{2.9.2}
\end{eqnarray}
where integers $m$ and $n$ enumerate the knot number of $\varphi (x)$ and
$w(x)$ functions respectively. According to this we shall denote the
boundary value $\varphi (0)$ and parameter $g$ in the following manner:
$\varphi _{mn}^*$ and $g_{mn}^*$. The result of numerical calculations
on Fig.1,2 are displayed ($w_1=0.1$).
\par
The asymptotic behaviour of the $\varphi _{mn}(x)$ and $w_{mn}(x)$
functions as in (\ref{2.9.1})-(\ref{2.9.2}) results in that the energy
density of this fields drop to zero as exponent on the infinity and this
means that this tube has the finite energy per unit. It is easy
to show that a flux of colour "magnetic" field $H_z$  across the plane
$z=const$ is finite.
\par
Thus we can to speak that the Yang - Mills - Higgs theory have
the tube solution if the Higgs field have the negative energy density.
It is notice that this solutions are not topological nontrivial
thread. Ignoring the transversal size of obtained tube we receive
the rested boson string with finite linear energy density.

\end{document}